\documentclass[fleqn,10pt,twocolumn, final]{NPGmain}
\usepackage[utf8]{inputenc}
\usepackage[T1]{fontenc}
\usepackage{graphicx}
\usepackage{subcaption} 
\usepackage{siunitx}
\usepackage{float}
\usepackage{stfloats}
\usepackage{authblk}
\usepackage[switch]{lineno}
\usepackage{lettrine}
\usepackage{soul}
\usepackage{pdfpages}
\usepackage{orcidlink}
\usepackage{anyfontsize}
\usepackage{siunitx}

\usepackage{braket}
\soulregister\cite7
\newcommand{\ketbra}[2]{\ket{#1}\bra{#2}}
\soulregister\ref7 
\begin{document}
\nolinenumbers 
\newcommand{\marco}[1]{\textcolor{red}{#1}}

\title{\LARGE Time-resolved certification of frequency-bin entanglement over multi-mode channels}
\author[1,*]{St\'ephane Vinet\orcidlink{0000-0002-2522-3193}}
\author[2]{Marco Clementi\orcidlink{0000-0003-4034-4337}}
\author[2]{Marcello Bacchi\orcidlink{0009-0003-8663-6754}}
\author[1]{Yujie Zhang\orcidlink{0000-0002-7858-7476 }}
\author[3]{Massimo Giacomin\orcidlink{0009-0001-1966-0549}}
\author[1]{Luke Neal\orcidlink{0000-0002-5766-1462 }}
\author[3]{Paolo Villoresi\orcidlink{0000-0002-7977-015X}}
\author[2]{Matteo Galli}
\author[4]{Daniele Bajoni\orcidlink{0000-0001-6506-8485}}
\author[1,5]{Thomas Jennewein}

\affil[1]{Institute for Quantum Computing and Department of Physics \& Astronomy, University of Waterloo, 200 University Ave W, Waterloo, Ontario N2L 3G1, Canada}
\affil[2]{  Dipartimento di Fisica “A. Volta”, Università di Pavia, Via A. Bassi 6, 27100 Pavia, Italy}
\affil[3]{Dipartimento di Ingegneria dell’Informazione and Quantum Technologies Research Center, Università degli Studi di Padova, Via Gradenigo 6B, 35131 Padua, Italy}
\affil[4]{Dipartimento di Ingegneria Industriale e dell'Informazione, Università di Pavia, Via Ferrata 5, 27100 Pavia, Italy}
\affil[5]{Department of Physics, Simon Fraser University, 8888 University Dr W, Burnaby, British Columbia V5A 1S6, Canada}

\affil[*]{ Corresponding author. Email: svinet@uwaterloo.ca}

\begin{abstract}
\noindent\textbf{Abstract} \\
Frequency-bin entangled photons can be efficiently produced on-chip which offers a scalable, robust and low-footprint platform for quantum communication, particularly well-suited for resource-constrained settings such as mobile or satellite-based systems.
However, analyzing such entangled states typically requires active and lossy components, limiting scalability and multi-mode compatibility. We demonstrate a novel technique for processing frequency-encoded photons using linear interferometry and time-resolved detection. Our approach is fully passive and compatible with spatially multi-mode light, making it suitable for free-space and satellite-to-ground applications. 
As a proof-of-concept, we utilize frequency-bin entangled photons generated from a high-brightness multi-resonator source integrated on-chip to show the ability to perform arbitrary projective measurements  over both single- and multi-mode channels. 
We report the first measurement of the joint temporal intensity between frequency-bin entangled photons, which allows us to certify entanglement by violating the Clauser–Horne–Shimony–Holt (CHSH) inequality, with a measured value of $|S|=2.32\pm0.05$ over multi-mode fiber. 
By combining time-resolved detection with energy-correlation measurements, we perform full quantum state tomography, yielding a state fidelity of up to $91\%$. We further assess our ability to produce non-classical states via a violation of time-energy entropic uncertainty relations and investigate the feasibility of a quantum key distribution protocol.
Our work establishes a resource-efficient and scalable approach toward the deployment of robust frequency-bin entanglement over free-space and satellite-based links.

\end{abstract}

\twocolumn
\captionsetup[figure]{labelfont={bf},name={Fig.},labelsep=none}
\maketitle
\section*{Introduction}
\noindent
Frequency-bin-entangled qubits \cite{Olislager2010, Lu:23} are versatile quantum information resources, well-suited for applications in quantum communication, computation, and sensing. They can be readily generated in chip-scale microresonators \cite{Reimer2016, clementi_programmable_2023}, transmitted over single-mode fiber, and support dense wavelength-division multiplexing (DWDM) thus enabling parallelized quantum communication. Moreover, they naturally scale to high-dimensional states, or \textit{qudits}\cite{kues_-chip_2017, borghi2023reconfigurablesiliconphotonicschip}, and their low size, weight, and power (SWaP) requirements make them excellent candidates for quantum communication systems in resource-constrained environments, such as satellite payloads \cite{Vinet:25, hyperspace_cordis_2025}.
However, the ability to efficiently manipulate and analyze frequency-bins remains limited, as traditional techniques rely on active non-linear optics to modify photon frequencies, e.g. through sum- and difference-frequency generation in $\chi^{(2)}$ media, which requires stringent phase-matching conditions and high pump powers that restrict both flexibility and scalability \cite{kobayashi_frequency-domain_2016,Serino:25,v24q-sl6n}. 
More recently, coherent control over frequency modes has been demonstrated using electro-optic modulators in combination with pulse-shaping techniques \cite{Lukens:17, kues_-chip_2017,PhysRevLett.120.030502}. 
While more practical, these systems inherently suffer losses due to unwanted sidebands, especially when relying on high-order modulation beyond the desired frequency range, and require complex, expensive, and typically lossy components to operate.
Furthermore, scaling these methods to manipulate multiple frequency modes requires intricate setups involving cascaded components, whose configurations must be extensively optimized \cite{lu_bayesian_2022}.  
Additionally, these approaches rely on single-spatial-mode operation, which limits their applicability for free-space channels, unless adaptive optics are employed. 
Expanding frequency-bin compatibility to ground- and satellite-based free-space channels is crucial for enabling truly global quantum networks, where atmospheric turbulence and pointing instability necessitate robust multi-mode operation to maintain high-fidelity quantum communication. 
The disparity between the relative ease of state generation and the complexity of manipulation underscores the need for alternative strategies in frequency-domain quantum information processing. 

In this work, we propose and demonstrate a novel technique based on passive photonic components and time-resolved detection \cite{PhysRevLett.112.103602,halder_entangling_2007}. Using the fact that the beating signal, determined by the frequency separation $\Delta\omega$, is within the timing resolution of the detection system $(\delta t \ll \frac{2\pi}{\Delta \omega})$, quantum interference in the two-photon state can be measured directly \cite{Cui2020, Guo:17}. 
The joint temporal intensity (JTI) effectively captures the temporal correlations between the signal and idler photons and encodes their frequency-bin superposition state.
\begin{figure*}[!t]
    \centering
    \includegraphics[width=\linewidth]{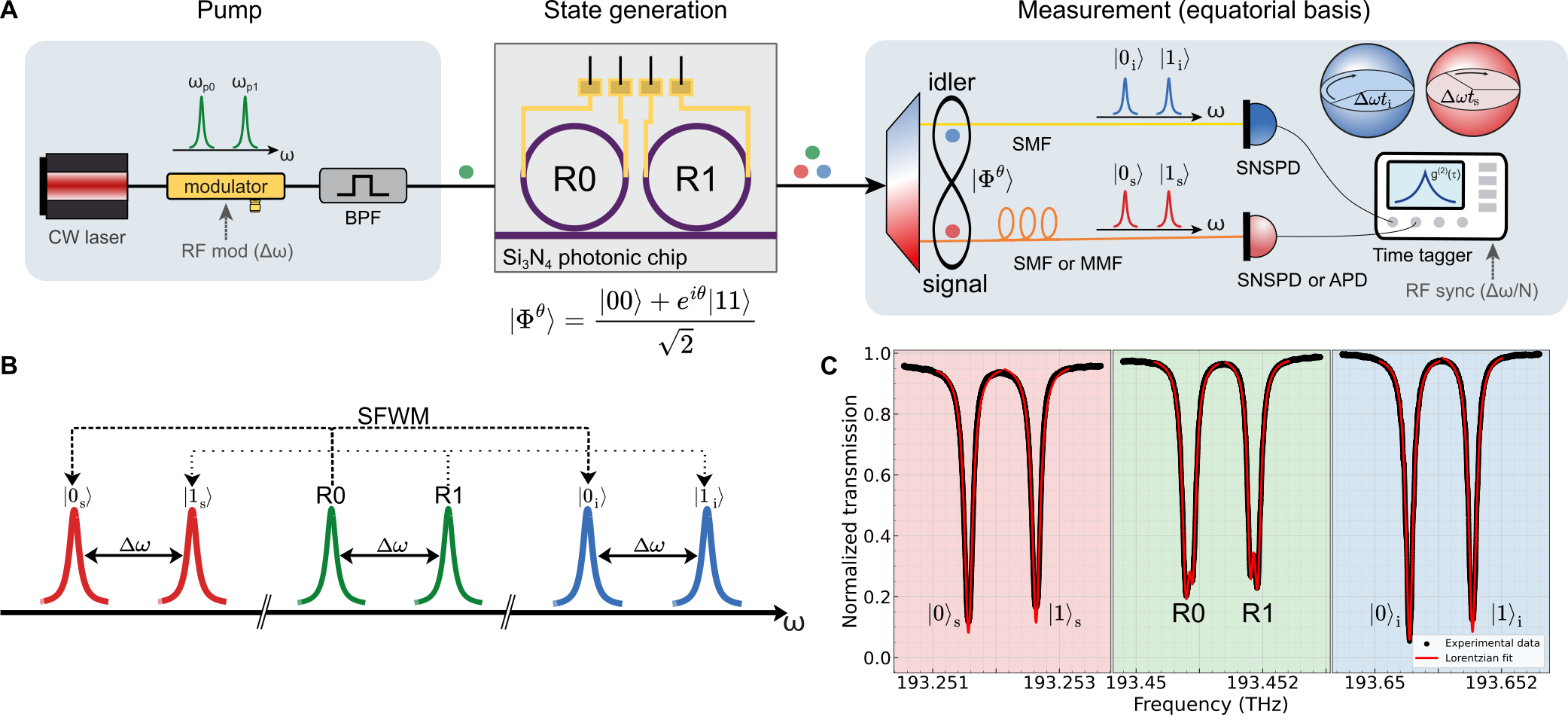}
    \caption{\small \textbf{. State generation and time-resolved detection.} 
    \textbf{(a)} Experimental layout for state generation and equatorial basis measurement. 
    \textbf{(b)} Schematic of the mechanism for the generation of the entangled state: the idler (blue) and signal (red) frequency bins are generated from the bichromatic pump (green) via spontaneous four-wave mixing (SFWM).
    All the resonances (as well as the pump fields) are spaced by $\Delta\omega/2\pi=\SI{820}{\mega\hertz}$.
    \textbf{(c)} Experimental transmission spectra and Lorentzian fit of the device at the signal, pump and idler frequencies. 
    Pump spectra show split resonances due to Rayleigh backscattering (see Methods) and are fitted using a two-Lorentzian model. 
    RF: radio-frequency; BPF: band-pass filter; SMF: single-mode fiber; MMF: multi-mode fiber; SNSPD: superconducting nanowire single photon detector; APD: avalanche photodiode; SFWM: spontaneous four-wave mixing.
 }
    \label{fig:timeresolved}
\end{figure*}
We hereby show that arbitrary projective measurements in the equatorial plane of the Bloch sphere can be implemented by post-selecting on various relative time delays.
Furthermore, beyond the equatorial basis, any projective measurement can be realized using high temporal resolution detectors in combination with linear interferometry. In particular, we demonstrate projections onto the computational basis (i.e., the eigenvectors of the Pauli operator $\sigma_Z$) using standard demultiplexing techniques with an unbalanced Mach–Zehnder interferometer (MZI) \cite{Vinet:25}.
Transmitting over multi-mode channels, however, creates spatial distortions that compromise distinguishability in unbalanced interferometric systems \cite{Jin:19,PhysRevLett.116.253601}. 
Spatial filtering can correct the path distinguishability but at a substantial throughput efficiency cost. 
To mitigate this issue, we here employ field-widened interferometers \cite{Shepherd:85,Liu:12}, which preserve path indistinguishability without the need for modal filtering or any adaptive optics. 
This approach is purely passive, requires minimal resources, and naturally extends to higher dimensions. 

\section*{Results}
\subsubsection*{Photon pair generation}
A conceptual schematic of the frequency-bin entangled photon source and equatorial basis measurement apparatus is shown in Fig.~\ref{fig:timeresolved}\textcolor{blue}{a}.
The source design is based on a multiple-resonator approach\cite{Liscidini2019, clementi_programmable_2023, borghi2023reconfigurablesiliconphotonicschip, Tagliavacche2025, Pang2025}, originally developed to simultaneously provide high brightness and a moderate frequency-bin spacing, accessible by commercial modulators.
It consists of a sequence of two nominally identical microring resonators, each associated with a state of the computational basis and hence labelled, respectively, R0 and R1.
The light from a tunable and continuous wave (CW) pump laser (\textit{Santec TSL-570}) at 1550.06 nm is sent through an electro-optic phase modulator (\textit{Exail MPZ-LN-10}) driven by a sinusoidal RF signal (\textit{AnaPico APMS20G}) at a frequency  $\Delta\omega/2\pi=\SI{820}{\mega\hertz}$. 
The resulting bichromatic field at the modulator's output (neglecting the sidebands not used in the experiment) is then coupled to a silicon nitride (Si$_3$N$_4$) photonic chip (insertion loss: \SI{3}{\deci\bel}/facet).
Here, each frequency component excites a respective high-$Q$ microring resonance previously, set at $\omega_\mathrm{p_0}$ and $\omega_\mathrm{p_1}$ by thermo-optic tuning, provided by integrated resistive heaters placed above each ring, with mode spacing adjusted to match $\Delta\omega=|\omega_\mathrm{p_1}-\omega_\mathrm{p_0}|$. 
Photon pair generation from spontaneous four-wave mixing (Fig.~\ref{fig:timeresolved}\textcolor{blue}{b}) occurs in both rings in a coherent superposition, where each ring, after coarse-graining, 
contributes a \emph{single} pair of signal-idler frequency
modes, encoding the computational basis states $\ket{0_\mathrm{s,i}}$ and $\ket{1_\mathrm{s,i}}$.
The process is described by the two-photon creation operator:  
\begin{equation}
\hat B_{n} = \iint d\omega_\mathrm{s}\, d\omega_\mathrm{i}\,
\phi_n(\omega_\mathrm{s})\psi_n(\omega_\mathrm{i})\delta(\omega_\mathrm{s} + \omega_\mathrm{i} - 2\omega_{\mathrm{p}_n})\hat a^{\dagger}_\mathrm{s}(\omega_\mathrm{s}) \hat a^{\dagger}_\mathrm{i}(\omega_\mathrm{i})
\end{equation}
obtained from the continuous frequency mode creation operators $\hat a^{\dagger}_\mathrm{s}(\omega)$ and $\hat a^{\dagger}_\mathrm{i}(\omega)$, where the envelope functions $\phi_n(\omega)=
  \sqrt{\frac{\gamma}{2\pi}}\,
  \frac{1}{\omega-\omega_{\mathrm{s}_n}+i\gamma/2} $ and $
\psi_n(\omega)=
  \sqrt{\frac{\gamma}{2\pi}}\,
  \frac{1}{\omega-\omega_{\mathrm{i}_n}+i\gamma/2}
$ correspond to Lorentzian cavity modes of linewidth $\gamma$, centered at $\omega_{\mathrm{s}_n}$ and $\omega_{\mathrm{i}_n}$, respectively, and $n=0,1$. 

The modulation index is adjusted to obtain sidebands that equalize the probabilities of generating a photon pair in each ring. 
Assuming the pump power is sufficiently low to suppress multi-pair emission, the generated frequency-bin state can then be approximated by the Bell state given in Eq.~\eqref{eq:Bell}. 
More specifically, under the approximation $\Delta \omega\gg \gamma$, each envelope function defines an orthogonal mode:
$\hat a^{\dagger}_{\mathrm{s},n}=\int d\omega_\mathrm{s}\phi_n(\omega_\mathrm{s})\hat a^{\dagger}_\mathrm{s}(\omega_s), \hat a^{\dagger}_{\mathrm{i},m}=\int d\omega_\mathrm{i}\psi_m(\omega_\mathrm{i})\hat a^{\dagger}_\mathrm{i}(\omega_i),$
such that $[\hat a_{\nu,n},\hat a^{\dagger}_{\mu,m}]=\delta_{\nu\mu}\delta_{nm}$.
This condition is verified by the use of high-$Q$ resonators, here around $1.2\times10^6$ for the signal resonances and $1.3\times10^6$ for the idler resonances at nearly critical coupling (Fig.~\ref{fig:timeresolved}\textcolor{blue}{c}), corresponding to an intrinsic $Q$ factor of approximately $3.5\times10^6$.
In the Heisenberg picture, all dynamics sit in the field operators and the two-photon state remains time-independent: 
\begin{equation}\label{eq:Bell}
|\Phi^\theta\rangle\simeq
\frac{\hat B_{0}+e^{i\theta}\hat B_{1}}{\sqrt{2}}\ket{\rm vac}\simeq
\frac{\ket{0_\mathrm{s}0_\mathrm{i}}+e^{i\theta}\ket{1_\mathrm{s}1_\mathrm{i}}}{\sqrt2}
\end{equation}
where $\ket{\rm vac}$ is the vacuum state and the phase $\theta$ is given by (twice) the relative phase between the pump fields driving R0 and R1.
With this procedure we have constructed “piece-by-piece” a frequency-bin entangled state (specifically, a generic $\Phi$-type Bell state), represented by the coherent superposition of two coherent two-photon squeezers, with a strategy analogous to the one described in Ref.\cite{clementi_programmable_2023}.
Note that the bin separation $\Delta\omega/2\pi=\SI{820}{\mega\hertz}$ is very low compared to the one typically achieved in single-resonator systems, which would require a resonator length of the order of tens of centimeters.
Instead, our approach enables us to preserve a high finesse, and therefore low power requirements (i.e. high brightness), and a low footprint. We stress that the coherence between the terms $\ket{0_\mathrm{s}0_\mathrm{i}}$ and $\ket{1_\mathrm{s}1_\mathrm{i}}$ is inherited from the coherence of the bichromatic field emitted by the phase modulator. As a consequence, the coherence time of the generated Bell state corresponds to that of the pump laser ($\sim \SI{1}{\milli\second}$). 
\subsubsection*{Time-resolved detection}
A DWDM filter separates the idler (1548.5 nm) and the signal (1551.7 nm) photons, which are then routed to high temporal resolution single-photon detectors.
A coincident detection event between the signal arm at time~$t_\mathrm{s}$ and the idler arm at time~$t_\mathrm{i}$ can be described by the time-dependent projection-valued measure (PVM) $\hat\Pi(t_\mathrm{s},t_\mathrm{i})=\ket{P_\mathrm{s,i}}\bra{P_\mathrm{s,i}}$,  where:
\begin{equation}
\ket{P_\mathrm{s,i}(t_\mathrm{s},t_\mathrm{i})}
= \frac{1}{2}
\left(\ket{0_\mathrm{s}}+e^{i\Delta\omega t_\mathrm{s}}\ket{1_\mathrm{s}}\right)
\otimes 
\left(\ket{0_\mathrm{i}}+e^{i\Delta\omega t_\mathrm{i}}\ket{1_\mathrm{i}}\right). \label{eq:P_proj}
\end{equation}
\\
The expectation value of $\hat\Pi$ on the state $\ket{\Phi^\theta}$ provides a representation on the continuous temporal basis, equivalent to the Glauber second-order correlation function\cite{Loudon2000} in the low squeezing regime.
In the framework of a spontaneous parametric source, this is also known as the joint temporal intensity \cite{Kuzucu2008, Borghi2024}, here taking the form:
\begin{align}
J(t_\mathrm{s}, t_\mathrm{i})&=|\langle P_\mathrm{s,i}(t_\mathrm{s},t_\mathrm{i})\ket{\Phi^\theta}|^2 \notag \\
&\propto
e^{-\gamma|t_\mathrm{s}-t_\mathrm{i}|}\,
\Bigl[1+\cos\!\bigl(\Delta\omega(t_\mathrm{s}+t_\mathrm{i})+\theta\bigr)\Bigr].
\label{eq:jta}
\end{align}
Eq.~\eqref{eq:jta} exhibits a clear dependence on the signal and idler arrival times $t_\mathrm{s}, t_\mathrm{i}$ and we interpret it as follows: a measurement performed at time $t_\mathrm{s}$ ($t_\mathrm{i}$) corresponds to a projective measurement upon a signal (idler) qubit state located at the equator of the Bloch sphere with azimuthal phase $\Delta\omega \, t_\mathrm{s}$ ($\Delta\omega \, t_\mathrm{i}$), as schematically shown in Fig.~\ref{fig:timeresolved}\textcolor{blue}{a}.
This interpretation stems evidently from the definition of the projector (Eq.~\eqref{eq:P_proj}) which in turn is legitimized by the choice of the frequency-bin basis.
The JTI can thus be formally interpreted as a result of the PVM defined above, effectively operating a time-resolved detection on the basis of the frequency bins. 
As an example, a set of projections for the $\sigma_X\sigma_X$ basis is depicted on the theoretical JTI shown in Fig.~\ref{fig:equatorialdata}\textcolor{blue}{a}, with the corresponding Bloch sphere projection illustrated in Fig.~\ref{fig:equatorialdata}\textcolor{blue}{c}. The theoretical JTI features an oscillating term as a function of $t_\mathrm{s}+t_\mathrm{i}$, which can be interpreted as quantum interference analogous to that observed in a Bell nonlocality experiment\cite{Freedman1972} or in Franson's scheme\cite{Franson1989}, and can be leveraged to certify entanglement, as discussed in the next sections. 
 The JTI also exhibits a decay term related to the relative time delay $|t_\mathrm{s}-t_\mathrm{i}|$, which is associated with the resonator ringdown time $1/\gamma$. 
 This decay results in an effective measurement described by a lossy projector with time-dependent efficiency \(\eta(t_\mathrm{s}, t_\mathrm{i}) = e^{-\gamma |t_\mathrm{s} - t_\mathrm{i}|}\).
 Note that this efficiency is close to unity as long as the resonances are well-resolved $\Delta\omega\gg\gamma$, a condition already satisfied by the near-orthogonality of the frequency-bin states. 

\subsubsection*{Equatorial basis: joint temporal intensity measurement}
\label{sec:correlation}
\begin{figure*}[h!]
    \centering
    \includegraphics[width=\linewidth]{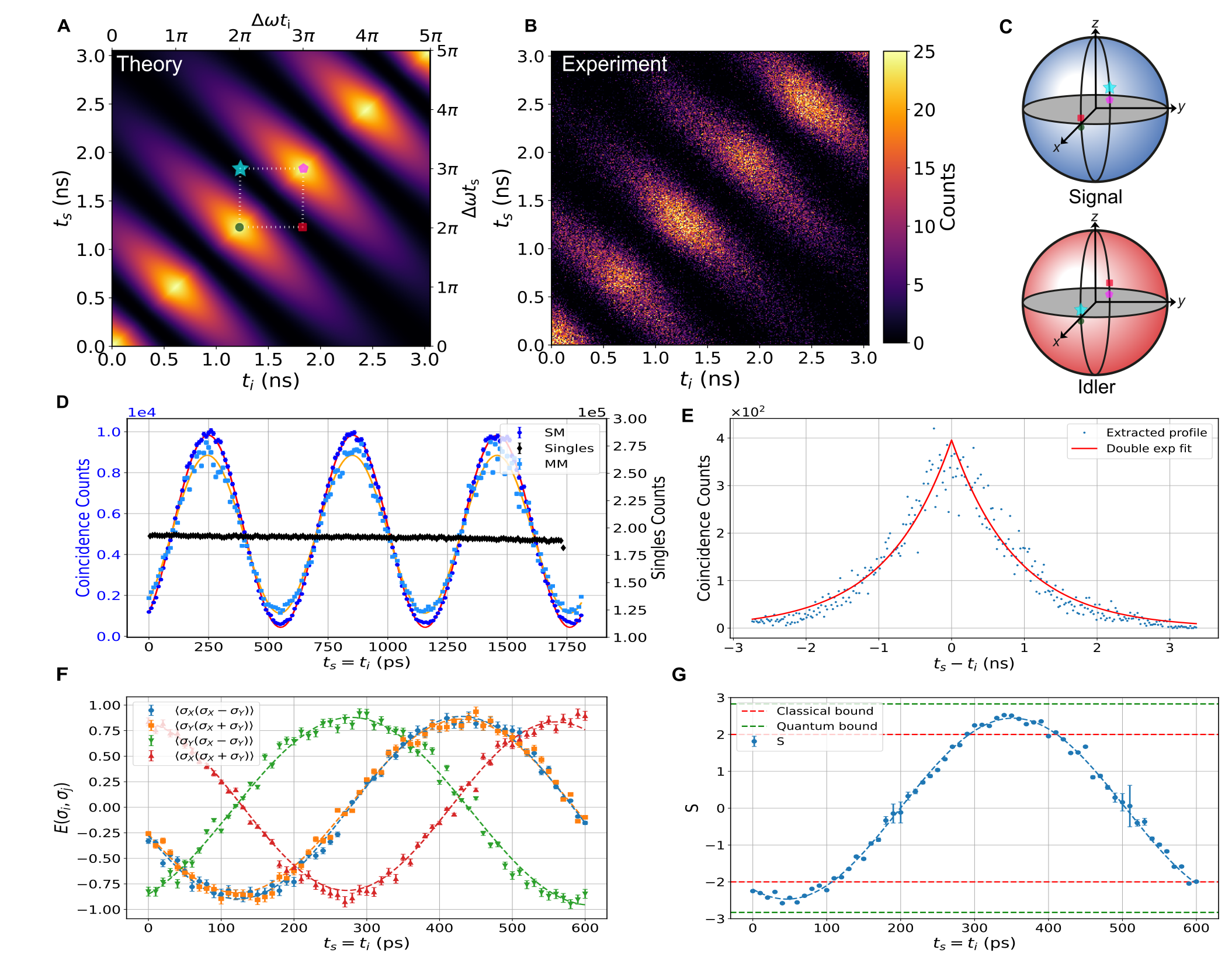}
    \caption{\small \textbf{. Equatorial basis measurements.} 
    \textbf{(a)} Theoretical JTI for ringdown time $1/\gamma=0.90$ $\rm{ns}$, the counts were scaled to match the experimental data. 
    \textbf{(b)} Measured JTI over single-mode fiber, with a fitted ringdown time $1/\gamma_\mathrm{fit}=0.90(4)$ $\rm{ns}$.
    \textbf{(c)} Visual representation of the equatorial basis projection for signal and idler qubits, $\sigma_X\sigma_X$ projections on the Bloch spheres are pictorially represented as colored points on the JTI. 
    \textbf{(d)} Diagonal profile of the measured JTI, a biphoton temporal beating can be observed in the coincidences whereas the singles exhibit no oscillation highlighting the nonlocal nature of the correlations. The single-mode and multi-mode signal propagation scenarios are denoted respectively as SM and MM.
    \textbf{(e)} Antidiagonal profile of measured JTI, the cavity lifetime leads to a double exponential decay proportional to $\gamma$. 
    \textbf{(f)} CHSH expectation values $\langle A_\mathrm{i} B_j \rangle,$ $i,j \in \{0,1\}$ as a function of the absolute time $(t_\mathrm{s}=t_\mathrm{i}) \mod T_\mathrm{b}$, and resulting \textbf{(g)} CHSH $S$ parameter (see Eq.~\eqref{eq:CHSH}). 
    The time axis spans 0 to 610 ps, corresponding to one beat period $T_\mathrm{b}=(\pi/\Delta\omega)$ and $i,j$ denote different azimuthal angles on the equator of the Bloch sphere. 
    The integration time for the equatorial basis measurement was 600 seconds. }
    \label{fig:equatorialdata}
\end{figure*}
We implement the equatorial measurement described above over both single- (\textit{Thorlabs SMF-28}) and multi-mode (\textit{Thorlabs GIF625}) fiber, with signal detection performed using a superconducting nanowire single-photon detector (SNSPD) (\textit{PhotonSpot}) in the single-mode case, and an InGaAs avalanche photodiode (APD) (\textit{MPD PDM-IR}) in the multi-mode case. The specifications for both detectors are provided in Table~\ref{tab:detector}.
\begin{table}[ht!]
    \centering
    \begin{tabular}{|c|c| c|c|}
    \hline
        Detector & DCR (cps) & Dead time (ns) & $\delta t$ (ps)\\
        \hline
         SNSPD & 300 & 70 & 60 \\
         InGaAs APD& 17330 &$10^4$& 72\\ 
         \hline
    \end{tabular}
    \caption{Specifications for the detectors used in the experiment. DCR denotes the dark count rate and $\delta t$ corresponds to the FWHM timing resolution. The InGaAs APD was operated at $6$V excess bias, in an externally gated mode, triggered by the SNSPD.}
    \label{tab:detector}
\end{table}
In both scenarios, idler photons are coupled into a single-mode fiber and detected on a SNSPD channel. To reduce the dark count rate, the InGaAs APD was operated in an externally gated mode, triggered by the idler SNSPD. 
A marker pulse at $\Delta\omega/40\pi=\SI{41}{\mega\hertz}$ provides a time reference to the time-tagging unit (\textit{Swabian instruments Time Tagger X}). 
Leveraging detectors with timing resolutions $\delta t\leq 60$ ps (full width at half maximum jitter), we measured, to the best of our knowledge, the first JTI for frequency-bin entangled photons in Fig.~\ref{fig:equatorialdata}\textcolor{blue}{b}. 
The measured JTI shows excellent agreement with the theoretical expectation shown in Fig.~\ref{fig:equatorialdata}\textcolor{blue}{a}.
The oscillatory term $\cos\!\bigl(\Delta\omega(t_\mathrm{s}+t_\mathrm{i})+\theta\bigr)$ along the diagonal of the JTI, shown in Fig.~\ref{fig:equatorialdata}\textcolor{blue}{d}, has a beating visibility of 91.9(9)$\%$, calculated by integrating along the anti-diagonals with a range $|t_\mathrm{s}-t_\mathrm{i}|\leq800$ ps. 
Its antidiagonal profile, shown in Fig.~\ref{fig:equatorialdata}\textcolor{blue}{e} and calculated by integrating along the diagonals, follows the exponential envelope $e^{-\gamma|t_\mathrm{s}-t_\mathrm{i}|}$, with ringdown time $1/\gamma=0.90(4)$ $\rm{ns}$. 
Importantly, the temporal oscillation appears only in the coincidence counts, and should therefore be regarded as a property of the two-photon state, while the single counts, related to the marginal states, remain flat, thus highlighting the two-photon, non-classical nature of the interference.
We stress that this cannot be attributed to the sole temporal oscillation of the pump field driving the system, as would be the presence of unresolved resonances ($\gamma\gg\Delta\omega$) or in a non-resonant system, such as a waveguide.

For a fixed relative time delay, 
$t_\mathrm{s}-t_\mathrm{i}$, the two-photon correlation (as shown in Fig.~\ref{fig:equatorialdata}\textcolor{blue}{f} for $t_\mathrm{s}-t_\mathrm{i}=0$) is analogous to that obtained in polarization entanglement where the joint detection time, $(t_\mathrm{s}=t_\mathrm{i}) \mod T_\mathrm{b}$, where $T_\mathrm{b}=\frac{\pi}{\Delta\omega}$ is the biphoton quantum beat period \cite{Guo:17}, defines the relative phase of the Bell-basis projection. 
Alternatively, the phase setting can also be adjusted by tuning the relative arrival time of the signal and idler photons with respect to the phase modulation signal (see Supplementary Note 1). 
We can use the correlations between the detection times to test for quantum nonlocality. 
In particular, we consider the  Clauser–Horne–Shimony–Holt (CHSH) inequality \cite{PhysRevLett.23.880}:
\begin{equation}\label{eq:CHSH}
    S=|\langle A_0B_0\rangle +\langle A_0 B_1\rangle +\langle A_1B_0\rangle -\langle A_1 B_1\rangle|\leq 2.
\end{equation}
Figs.~\ref{fig:equatorialdata}\textcolor{blue}{f}-\ref{fig:equatorialdata}\textcolor{blue}{g} show the Bell correlations $\langle A_iB_j\rangle$ ($i,j \in \{0,1\}$), and the CHSH parameter $S$ as a function of detection time $t_\mathrm{s}=t_\mathrm{i}$.
Note that we choose the measurement operators $A_0=\sigma_X, A_1=\sigma_Y, B_0=\frac{1}{\sqrt{2}}(\sigma_X-\sigma_Y), B_1=\frac{1}{\sqrt{2}}(\sigma_X+\sigma_Y)$, i.e. all our measurements are in the equatorial plane of the Bloch sphere, as sketched in Fig.~\ref{fig:equatorialdata}\textcolor{blue}{a} and Fig.~\ref{fig:equatorialdata}\textcolor{blue}{c}, where the temporal sampling coordinates on the JTI and the associated projections on the Bloch sphere are shown respectively for the $\sigma_{X}\sigma_X$ measurement. More specifically, the green circle, pink pentagon, red square, and blue star, respectively, denote $\ketbra{++}{++}_{\mathrm{si}},\;
\ketbra{--}{--}_{\mathrm{si}},\;
\ketbra{+-}{+-}_{\mathrm{si}},\;
\ketbra{-+}{-+}_{\mathrm{si}}
$ projections. To optimize the timing resolution of the detection scheme, we first consider projections only onto SNSPD channels, and obtain $S=2.53\pm 0.03$, at $t_\mathrm{s}=t_\mathrm{i}=340$ ps, which violates the classical limit by $\sim18\sigma$. 
Note that due to the low coincidence-to-accidentals ratio (CAR) $\sim 4$, we applied a background subtraction to the raw coincidence counts to compensate for uncorrelated events and accurately estimate the CHSH parameter.
We also observe a violation of $\sim6\sigma$ over multi-mode fiber, with $S=2.32\pm 0.05$. 
The primary difference with the single-mode fiber case is the reduced statistics, due to the lower detection efficiency of the InGaAs APD (see Methods, Table~\ref{tab:losses}).
\subsubsection*{Field-widened interferometer}
\begin{figure*}[h!]
    \centering
    \includegraphics[width=\linewidth]{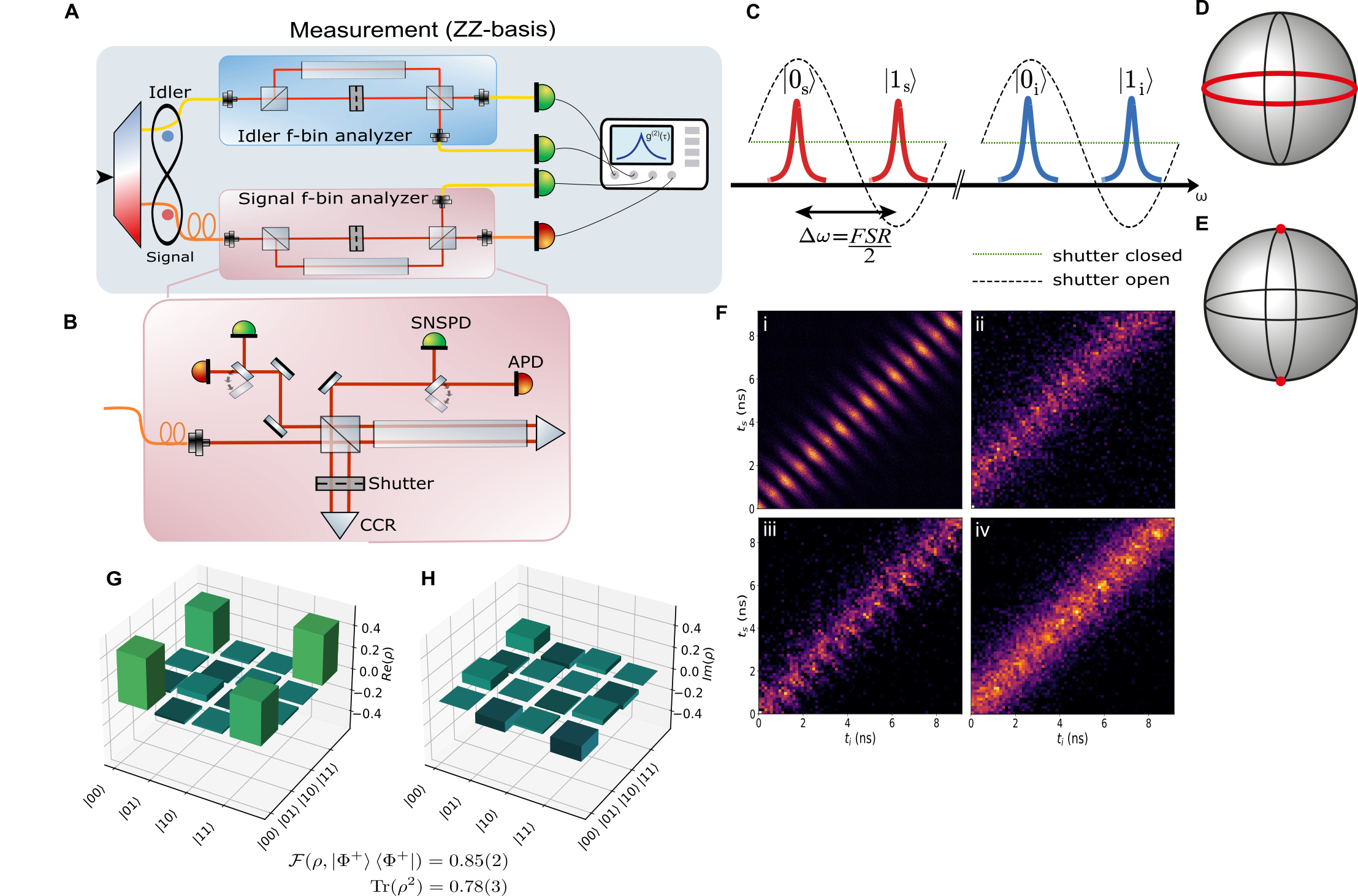}
    \caption{\small \textbf{. Interferometric detection scheme.} 
    \textbf{(a)} Experimental layout for the $\sigma_Z\sigma_Z$ measurement. 
    \textbf{(b)} A detailed schematic of the field-widened interferometer. 
    A \SI{10}{\centi\meter
    } N-BK7 glass rod in the long arm of the interferometer is used to compensate for spatial distortions of the optical mode of the frequency-bin qubit. 
    The basis choice is implemented via an optical shutter in the short arm of the interferometer. 
    Flip-mirrors at the interferometer's outputs enable coupling into single- and multi-mode fibers. 
    \textbf{(c)} Conceptual diagram of the interferometric frequency demultiplexing. 
    The signal (idler) resonances are spaced by half of the interferometer free spectral range (FSR), which acts as a narrowband filter. 
   \textbf{(d)} Visual representation of the projection for the closed (equatorial basis) and \textbf{(e)} open (computational basis) shutter  configurations. 
   \textbf{(f)} Joint temporal intensity (JTI) for the four physical measurement settings required for quantum state tomography:  (i) both photons in the equatorial basis, (ii) signal in the equatorial basis and idler in the $\sigma_Z$ basis, (iii) signal in the $\sigma_Z$ basis and idler in the equatorial basis, and (iv) both photons in the $\sigma_Z$ basis. 
   Quantum state tomography: reconstructed real \textbf{(g)} and imaginary \textbf{(h)} part of the density matrix of the state measured over multi-mode fiber. }
    \label{fig:ZZmeasurement}
\end{figure*}

To complement the equatorial basis measurements, we performed $\sigma_Z\sigma_Z$ measurements using Mach-Zehnder interferometers as depicted in Figs.~\ref{fig:ZZmeasurement}\textcolor{blue}{a-b}. 
By carefully tuning the path delay of the unbalanced Mach–Zehnder interferometer, we can effectively realize projections onto \(\{\ket{0_\mathrm{s(i)}}, \ket{1_\mathrm{s(i)}}\}\). The optical shutter in the short arm of each interferometer can implement an active basis choice between the computational and equatorial bases, as sketched in Figs.~\ref{fig:ZZmeasurement}\textcolor{blue}{c-e}. In practice, the interferometers were circumvented for the equatorial basis, as shown in Fig.~\ref{fig:timeresolved}, to avoid their insertion loss and increase the detected rate, which in turn allowed us to reduce the pump power and thereby improve the CAR.
In addition, cross-basis measurements were performed, in which one qubit was measured in the $\sigma_Z$ basis and the other in the equatorial basis, although this configuration is not explicitly shown.

While unbalanced interferometers are typically unsuitable for turbulent free-space channels, field-widened interferometers \cite{vanasse1977spectrometric, PhysRevA.97.043847,PhysRevLett.116.253601}, enhanced with specialized imaging optics or carefully calibrated refractive indices, can serve as multi-spatial mode frequency-bin quantum receivers by neutralizing angular variation-induced phase shifts and visibility degradation \cite{Shepherd:85,Liu:12}.
We here adopt the latter method: where the combination of the glass rod in the long arm and the retro-reflector (see Fig.~\ref{fig:ZZmeasurement}\textcolor{blue}{b}) creates a virtual mirror that appears closer to the interferometer’s beam splitter. 
Thus by carefully selecting the glass’ refractive index and length, we can match the distance from the beam splitter to this virtual mirror with the actual distance to the real mirror in the short arm, effectively balancing the interferometer. 
More specifically, consider a MZI with an input beam incident at an angle $\alpha$, we can approximate the optical path length difference, using Snell's law and Taylor's expansion for small $\alpha$, as:
\begin{equation}
\Delta L=
\sum_j n_{jl}L_{jl}-n_{js}L_{js}-
\frac{n_{0}^2\sin^2\alpha}{2}\left( \frac{L_{jl}}{n_{jl}}-\frac{L_{js}}{n_{js}}\right),
\end{equation}  
where $l,s$ respectively denote the long and short arms of the interferometer and $n_{js(l)}$ the different refractive indices \cite{Liu:12,PhysRevA.97.043847}. 
By appropriately choosing the refractive index of the optical elements $n_{js(l)}$ and lengths $L_{js(l)}$, we can eliminate to first order the angular dependence of the interferometer's optical path length difference.
On the other hand, for it to frequency demultiplex we require 
$\Delta L=\frac{\pi c}{\Delta \omega}$.
A multi-mode frequency qubit analyzer can be constructed by solving both of these conditions simultaneously. The specific interferometer configurations used in this experiment are described in the Methods section.
\subsubsection*{Computational basis: interferometric measurement}\label{sec:ZZ}
\begin{figure}
    \centering
    \includegraphics[width=\linewidth]{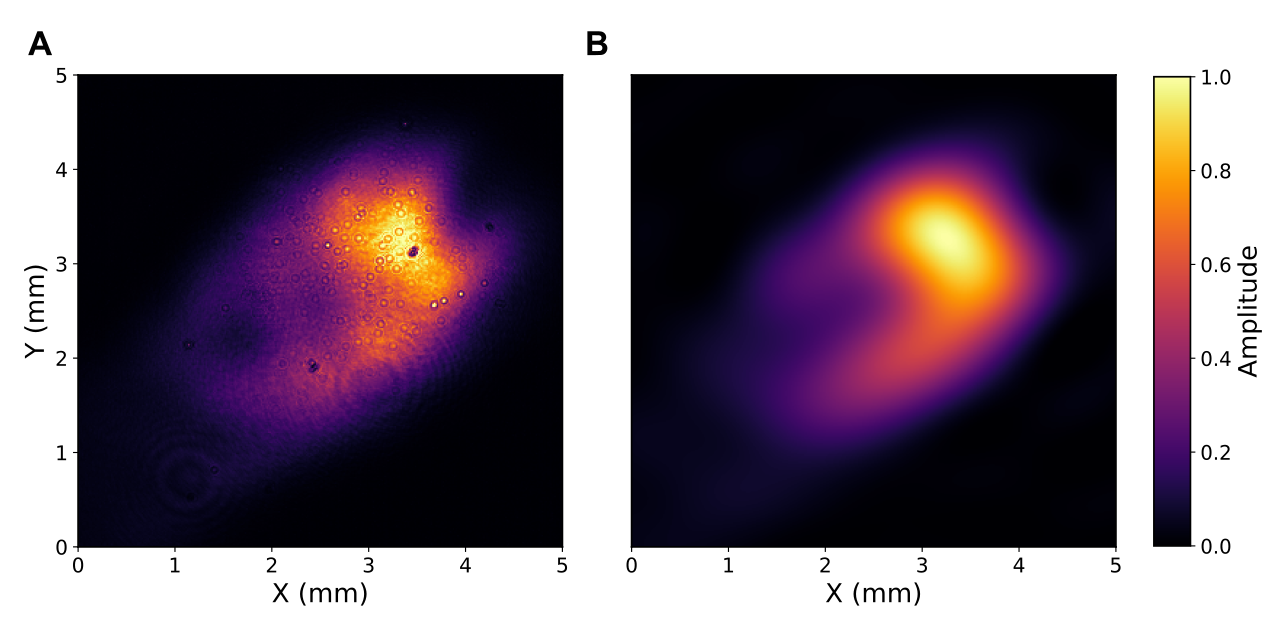}
   \caption{\small \textbf{. Point spread function.} 
    \textbf{(a)} PSF, imaged with a
beam-profiling camera (\textit{New Imaging Technologies WiDy SenS 640}), of the multi-mode beam in the analyzer, after propagation through a 5-meter long graded-index (GRIN) multi-mode
fiber (\textit{Thorlabs GIF625}). \textbf{(b)} Reconstruction of the PSF from modal decomposition using the first 30 Hermite-Gaussian modes.  }
    \label{fig:beam}
\end{figure}

\begin{table}
    \centering
    \begin{tabular}{|c|c| c|c| c|}
    \hline
    Projections & $\ketbra{00}{00}$ & $\ketbra{10}{10}$ &$\ketbra{01}{01}$&$\ketbra{11}{11}$\\
    \hline
      $C_{ij}$ & 264897 & 11011 & 25411&270383\\
      \hline
    \end{tabular}
    \caption{Computational basis ($\sigma_Z\sigma_Z$) measurement statistics. The counts $C_{ij}$, $i,j \in \{0,1\}$, corresponding to projections $\ketbra{ij}{ij}$,  were normalized by the channel efficiency. Note that the integration time was $60$ seconds and detections $C_{i1}$ were measured with the InGaAs APD. A detailed breakdown of the channel losses is given in Table~\ref{tab:losses} of the Methods section. }
    \label{tab:ZZcounts}
\end{table}
 Table~\ref{tab:ZZcounts} shows the counts for the $\sigma_Z\sigma_Z$-basis measurement, where the four detection outcomes $C_{ij}$, $i,j \in \{0,1\}$ correspond to the $\ket{ij}$ projections. 
 Despite significant spatial distortions (see Fig.~\ref{fig:beam}), introduced by the multi-mode fiber channel, the demultiplexing visibilities of the single- and multi-mode interferometers were, respectively, $92\%$ and $83\%$. 
 The optical throughput of the interferometers were limited by optical surface losses to $\sim5$ dB (see Table~\ref{tab:losses}).
 To limit experimental overhead, both interferometers operated without active phase stabilization, which constrained the measurement integration time. Further improvements in throughput and visibility are achievable via optimized spatial mode overlap, symmetric mode-matching optics, and enhanced thermal stability (see Methods), offering a straightforward path towards robust and scalable implementations.
\par To assess the performance of our time-resolved detection scheme, now capable to operate a tomographically complete set of measurements, we performed a full tomography of the generated state. 
A subset of the resulting joint temporal intensities, corresponding to the four physical measurement settings, is shown in Fig.~\ref{fig:ZZmeasurement}\textcolor{blue}{f} (the full set of JTIs is reported in Supplementary Note 2).
The temporal resolution of our detection scheme enables us to extract the full set of time correlations required for complete state reconstruction using only four physical measurement settings. Using a standard maximum likelihood estimation, we reconstruct a density matrix with an Uhlmann fidelity of $91(1)\%$ to the Bell state $\ket{\Phi^+}$ (that is, state $\ket{\Phi^\theta}$ with $\theta=0$), and a purity of $84(2)\%$ over the single-mode channel, whereas the density matrix for the multi-mode case, shown in Fig.~\ref{fig:ZZmeasurement}\textcolor{blue}{g-h}, yields a fidelity of $85(2)\%$ and a purity of $78(3)\%$.
A summary of the quantum state tomography fidelities and the CHSH values for the single- and multi-mode scenarios is presented in Table~\ref{tab:QST}.
\begin{table*}
    \centering
    \begin{tabular}{|c|c|c|c|c|}
    \hline
      Idler detector & Signal Detector  &$S$ & $\mathcal{F}(\rho,\ket{\Phi^+}\bra{\Phi^+}) $ & Tr$(\rho^2)$ \\
         \hline
        SNSPD & SNSPD & $2.53\pm0.03$ & $91\pm1\%$ & $84\pm2\%$ \\
        SNSPD & InGaAs APD & $2.32\pm 0.05$ & $85\pm2\%$ & $78\pm3\%$ \\
         \hline
    \end{tabular}
    \caption{Quantum state tomography fidelities, purities and CHSH $S$-parameters obtained using equatorial basis measurements performed over single-mode and multi-mode fiber. All remaining basis measurements were performed over multi-mode fiber as depicted in Fig.~\ref{fig:ZZmeasurement}.  }
    \label{tab:QST}
\end{table*}
\subsubsection*{Steering inequality and entropic uncertainty relation}
In the previous section, we showed that the JTI of
our two-photon state already certifies non-classicality through the violation of the CHSH inequality. 
In this section, we combine the $\sigma_Z$ measurement into the non-classicality certification. 

The task that we consider here relies on all three Pauli correlations $\sigma_X\otimes \sigma_X$, $\sigma_Y\otimes \sigma_Y$  and $\sigma_Z\otimes \sigma_Z$. The test can be formulated in terms of linear steering inequalities, defined in Eqs.~\eqref{eq:steering}-\eqref{eq:steering2},
and originally implemented for polarization-entangled photons \cite{saunders_experimental_2010}: 

\begin{align}
&\frac{1}{2}\bigg(|\langle\sigma_X\sigma_X\rangle|+|\langle\sigma_Z\sigma_Z\rangle| \bigg)\leq\frac{\sqrt{2}  }{2},  \label{eq:steering}\\
&\frac{1}{3}\bigg(|\langle\sigma_X\sigma_X\rangle|+|\langle\sigma_Y\sigma_Y\rangle|  +|\langle\sigma_Z\sigma_Z\rangle|\bigg)\leq\frac{\sqrt{3}  }{3}. 
\label{eq:steering2}
\end{align}
Violation of either bound certifies entanglement in the
one-sided-device-independent (steering) scenario \cite{uola2020}. 
The expectation values for the three bases considered in Eq.~\eqref{eq:steering2} are reported in Table~\ref{tab:steering}. 
In particular, we violate the classical bound in Eq.~\eqref{eq:steering2} (Eq.~\eqref{eq:steering}) by more than $27\sigma$ $(19\sigma)$ over single-mode fiber, and by $7\sigma$ $(5\sigma)$ over multi-mode fiber. 
Note that there is a slight discrepancy between the expectation values $\langle\sigma_X\sigma_X\rangle$ and $\langle\sigma_Y\sigma_Y\rangle$ for the equatorial basis. 
This asymmetry is also apparent as a parity dependent modulation in Fig.~\ref{fig:equatorialdata}\textcolor{blue}{b} and we attribute it to contributions from uncorrelated pump photons which introduce a sub-harmonic oscillation at $\Delta\omega(t_\mathrm{s}+t_\mathrm{i})/2$ (see Supplementary Note 1).
\begin{table*}
    \centering
    \begin{tabular}{|c|c|c|c|c|c|}
    \hline
      Idler detector & Signal Detector  &$\langle\sigma_X\sigma_X\rangle$ & $\langle\sigma_Y\sigma_Y\rangle$  & $\langle\sigma_Z\sigma_Z\rangle$ & $\frac{1}{3}\sum_{i=1}^3 \langle\sigma_\mathrm{i}\sigma_\mathrm{i}\rangle$
        \\
         \hline
        SNSPD & SNSPD & -0.912 & 0.839 & 0.873 & $0.875 \pm 0.011$ \\
        SNSPD & InGaAs APD & -0.880 & 0.831 & 0.873 & $0.861 \pm 0.039$ \\
         \hline
    \end{tabular}
    \caption{Steering inequality violation for the single-mode and multi-mode channels. Note that the $\langle\sigma_Z\sigma_Z\rangle$ was measured in both scenarios using the InGaAs APD detector. The counts were normalized using the channel efficiencies given in Table~\ref{tab:losses}. }
    \label{tab:steering}
\end{table*}

More interestingly, the same quantities can be evaluated through the discrete uncertainty relation studied in \cite{Maassen1988, Scheeloch2013}, which allows us to reveal the non-classical nature of our biphoton state through violation of the discrete time–energy uncertainty relations. 
Consider the discrete observables  $\sigma_X\otimes \sigma_X$, $\sigma_Y\otimes \sigma_Y$  and $\sigma_Z\otimes \sigma_Z$ measured in our experiment, the entropic uncertainty relations quantify the minimal joint uncertainty that two systems can have in different pairs of observables, which are of the form:
\begin{align}
H(\sigma_X^\mathrm{s}\mid\sigma_X^\mathrm{i})
+H(\sigma_Z^\mathrm{s}\mid\sigma_Z^\mathrm{i})
      &\ge 1 ,  \label{eq:entropic1}\\
H(\sigma_X^\mathrm{s}\mid\sigma_X^\mathrm{i})
+H(\sigma_Y^\mathrm{s}\mid\sigma_Y^\mathrm{i})
+H(\sigma_Z^\mathrm{s}\mid\sigma_Z^\mathrm{i})
      &\ge 2 ,\label{eq:entropic2}
\end{align}
where $H(\sigma_Z^\mathrm{s}\mid\sigma_Z^\mathrm{i})$ is the conditional Shannon entropy of the signal outcome given the idler outcome, which quantifies the correlation in the $\sigma_Z^\mathrm{s}\otimes \sigma_Z^\mathrm{i}$ measurements.  
More specifically, we denote the experimentally determined joint
probabilities as $\{p(a_\mathrm{s} b_\mathrm{i})\}_{a_\mathrm{s}, b_\mathrm{i}=0}^1$, the conditional Shannon entropy is then given as:
\begin{equation}
H(\sigma_Z^\mathrm{s}\mid\sigma_Z^\mathrm{i})=-\sum_{b_\mathrm{i}\in\{0,1\}} p(b_\mathrm{i}) \sum_{a_\mathrm{s}\in\{0,1\}}p(a_\mathrm{s}|b_\mathrm{i})\log_2p(a_\mathrm{s}|b_\mathrm{i})
\end{equation}
where $p(b_\mathrm{i})=\sum_{a_\mathrm{s}}p(a_\mathrm{s},b_\mathrm{i})$, $p(a_\mathrm{s}|b_\mathrm{i})=\frac{p(a_\mathrm{s},b_\mathrm{i})}{p(b_\mathrm{i})}$ and $p(a_\mathrm{s},b_\mathrm{i})$ is the probability of getting outcome $a_\mathrm{s},b_\mathrm{i}$ in the correlated Pauli measurement. 

These entropic inequalities are discrete-entropy versions of the
time–energy uncertainty relations introduced in
Refs.\cite{Reid1989,Cavalcanti2009}. 
In our setup, the $\sigma_Z$ basis corresponds to \textit{energy}
(i.e.\ frequency-bin) measurements, whereas the equatorial Pauli bases $\sigma_{X}$ and $\sigma_{Y}$ are realized via joint temporal measurements.  
Eqs.~\eqref{eq:entropic1}–\eqref{eq:entropic2} are therefore a
\emph{discrete version of the time–energy uncertainty relation}:
an entangled state can simultaneously render the conditional energy
entropy {\em and} the conditional time entropy almost zero, thus
violating the classical bound imposed on any local hidden state model \cite{Scheeloch2013}.
Here we summarize the experimental violation of the entropic uncertainty relation in Table~\ref{tab:entropy}.

\begin{table*}
    \centering
    \begin{tabular}{|c|c|c|c|c|c|}
    \hline
      Idler detector & Signal Detector  &$H(\sigma_X^\mathrm{s}\mid\sigma_X^\mathrm{i})$& $H(\sigma_Y^\mathrm{s}\mid\sigma_Y^\mathrm{i})$& $H(\sigma_Z^\mathrm{s}\mid\sigma_Z^\mathrm{i})$& $\sum_{j=1}^3 H(\sigma_j^\mathrm{s}\mid\sigma_j^\mathrm{i})$\\
         \hline
        SNSPD & SNSPD & 0.369 & 0.280 & 0.307  & $0.956 \pm 0.082$ \\
        SNSPD & InGaAs APD & 0.534 & 0.338 & 0.307 & $1.179 \pm 0.247$ \\
         \hline
    \end{tabular}
    \caption{Entropic uncertainty relation for the single-mode and multi-mode channels. Note that the $H(\sigma_Z^\mathrm{s}\mid\sigma_Z^\mathrm{i})$ was measured in both scenarios using the InGaAs APD detector.   }
    \label{tab:entropy}
\end{table*}

\subsubsection*{Quantum Key Distribution}
\begin{figure*}
    \centering
    \includegraphics[width=0.9\linewidth]{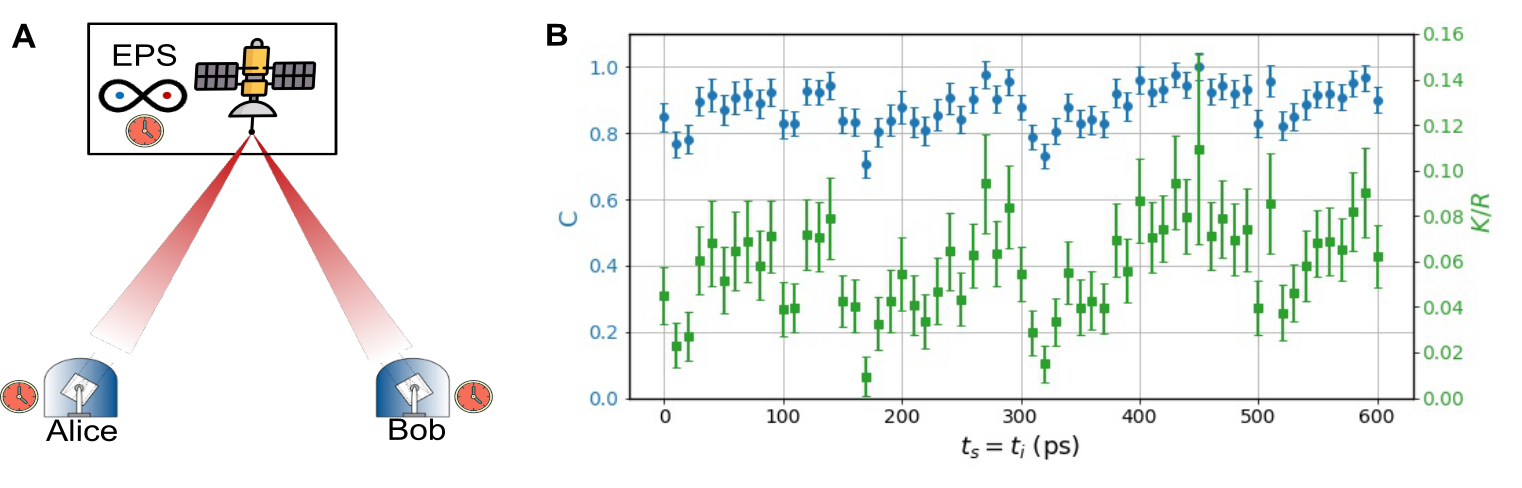}
    \caption{\small \textbf{. Quantum key distribution.} \textbf{(a)} Conceptual deployed scenario in a dual-downlink configuration, the entangled photon source (EPS) is onboard the satellite. \textbf{(b)} $C$-parameter ($C$) and normalized asymptotic key rate ($K/R$), calculated according to Eqs.~\eqref{eq:C}-\eqref{eq:SKR}. }
    \label{fig:applications}
\end{figure*}
Due to its compatibility with integrated photonic circuits, noise resilience \cite{cozzolino2019}, and favorable SWaP requirements \cite{Lu:23}, frequency-bin encoding is especially well suited for satellite-based quantum communication. 
There is a natural extension of this time-resolved frequency-bin detection scheme to quantum key distribution (QKD) \cite{pennacchietti_oscillating_2024}, in particular in the context of reference frame independent protocols (RFI-QKD).
While the equatorial basis can be used as an entanglement witness to estimate Eve's information, the \(\sigma_Z\)-basis is fixed and can be used for the key-map. Although our analysis assumes synchronized clocks at the users and the source (as depicted in Fig.~\ref{fig:applications}\textcolor{blue}{a}), the bichromatic pump could alternatively be used as a classical channel to provide a stable optical reference for clock synchronization between nodes, thus reducing the complexity of the system for practical satellite implementations. 
While the Doppler shift for low-Earth orbit satellites can induce an absolute frequency shift on the order of $\pm\SI{4}{\giga\hertz}$ \cite{PhysRevApplied.23.L021003}, this change is well-defined by the satellite's motion and can thus be tracked and compensated. Furthermore, the frequency separation $\Delta\omega/2\pi$ is only modified by $\sim \SI{21}{\kilo\hertz}$ - a negligible change compared to the frequency-bin bandwidth. A more detailed discussion of the feasibility of frequency-bin satellite quantum communication is provided in Ref.~\cite{Vinet:25}. To demonstrate feasibility, we consider the 6-state 4-state protocol defined in Ref.\cite{tannous_demonstration_2019, Tannous:25}, where the channel quality is evaluated using the following $C$-parameter:
\begin{equation}\label{eq:C}
    C=\sqrt{\langle\sigma_X\sigma_X\rangle^2+\langle\sigma_Y\sigma_X\rangle^2 },
\end{equation}
$C = 1$
indicates a maximally entangled state or perfect channel. 
The quantum bit error rate, meanwhile, is given by:
\begin{equation}
    \mathrm{QBER_{ZZ}}=\frac{1-\langle \sigma_Z\sigma_Z\rangle}{2}.
\end{equation}
Using the values calculated in Table~\ref{tab:steering}, we have $\rm{QBER_{ZZ}}=6.35\%$. We can then calculate an asymptotic key rate, $K$ according to:
\begin{equation}\label{eq:SKR}
    K\geq qR[1-fH_2(\mathrm{QBER}_{ZZ})-{H_2}(\mathrm{QBER_{\mathrm{Equatorial}}})],
\end{equation}
where $\mathrm{QBER}_{\mathrm{Equatorial}}=(1-C)/2$, $q$ is the basis reconciliation factor, $R$ the coincidence detection rate, $f$ the bidirection error correction efficiency, and $H_2(x)$ is the binary entropy function. 
    The main resulting figures of merit, the $C$-parameter and the secure key rate, are presented in Fig.~\ref{fig:applications}\textcolor{blue}{b}.
We observe that the $C$-parameter, which quantifies channel integrity, exceeds the threshold required for a positive secret key rate. 
The number of secret bits per coincidence detection at the asymptotic
limit is estimated to be $0.058(3)$ bits/coincidence relative to the theoretical maximum of $0.167$ bits/coincidence achievable with the 6-state 4-state protocol. 
We emphasize that these results are presented solely as a proof-of-concept, as the data have undergone post-processing including background subtraction. Nevertheless, these results indicate compatibility with practical quantum communication protocols, which may not be constrained by the extrinsic limitations of this demonstration, particularly the low CAR (see Methods).

\section*{Discussion}
We have demonstrated a novel approach to perform projective measurements on frequency-bin encoded photonic qubits using only interferometry and time-resolved detection, in a fully passive architecture, without relying on active modulation or spatial-mode filtering. From a fundamental perspective, frequency-bin encoded states are naturally suited for time-resolved measurements in the equatorial basis, as, in our detection scheme, the scan of the azimuthal phase is associated with  time-evolution.
This method enables arbitrary single-qubit projective measurements across the full Bloch sphere by combining interferometer phase tuning with time-resolved detection, which allows access to intermediate points between the poles and equator. Arbitrary qudit projective measurements can similarly be realized via additional linear interferometry, as outlined in Cui et al. \cite{Cui2020}. This latter approach has so far remained theoretical due to the technical difficulties of implementing frequency-bin encoded states with low frequency spacing. Here we address this issue, by adopting a two-resonator scheme, which overcomes the brightness-separation trade-off inherent to single-resonator sources \cite{clementi_programmable_2023}.
%
A time-resolved entanglement certification of frequency-bin qubits has also been proposed using cold atomic ensembles\cite{Guo:17}.
However, this scheme differs significantly with our and Cui's approach, relying on time-delay measurements, rather than absolute time evolution, in closer analogy with Ramsay interference\cite{Clemmen2016}.

In contrast, our approach provides the first theoretical and experimental demonstration of the JTI of a frequency-bin entangled state, which we use to visualize all the possible projections on the equatorial basis, effectively leveraging it as a projective measurement tool.
From the analysis of the JTI data, we retrieve important figures of merit of the source, such as the quantum interference visibility (91.9(9)\%), the biphoton coherence time ($1/\gamma=\SI{0.90(4)}{\nano\second}$), and the CHSH parameter ($S=2.53\pm 0.03$). Interestingly, Dekkers \textit{et al.} \cite{dekkers2025observinghighdimensionalbellinequality} recently demonstrated a conceptually related approach in the frequency domain, showing that the joint spectral intensity (JSI) is a phase-sensitive quantity containing the information required to violate the  Collins-Gisin-Linden-Massar-Popescu (CGLMP) inequality using time-bin entangled qudits \cite{PhysRevLett.88.040404}. This result underscores the understanding that Fourier-based frequency measurements provide direct access to the bases necessary for evaluating high-dimensional Bell inequalities. While arbitrary high-dimensional projectors would require additional interferometric stages or phase-shifting schemes, this time-resolved approach nevertheless enables straightforward certification of qudit entanglement. To exemplify this, a simulation of the JTI for a frequency-bin qutrit is provided in Supplementary Note 1.
The analysis capability is complemented by the use of interferometric filters, that extend the projective measurement capability beyond the equatorial subspace to span the entire two-qubit Hilbert space.
Access to a complete set of measurements enables the implementation of quantum state tomography, which we show to exhibit fidelities exceeding 91\% and purities above 84\%, comparable with those achieved in conventional frequency-bin analysis systems \cite{Imany:18, Morrison2022, clementi_programmable_2023}.
Importantly, our method is compatible with multi-mode propagation, as showcased by the experimental violations of the CHSH inequality and also of the quantum steering and entropic uncertainty relations, all performed over a multi-mode fiber link, with low performance degradation compared to the single-mode case.
These capabilities represent a significant step toward practical frequency-bin quantum networking, especially over free-space channels, where atmospheric turbulence results in spatial multimodality. 

As a proof-of-concept, we calculate the secure key rate for a RFI-QKD protocol using our measurement scheme. 
This demonstrates the potential of our approach for practical quantum cryptographic applications. 
Note that the implementation could be further simplified by restricting measurements to the equatorial basis of the Bloch sphere, which would reduce experimental complexity and receiver loss (here around $5$ dB) while still enabling secure key generation. The communication rate could be further increased through frequency multiplexing, by taking advantage of multiple free spectral ranges to encode several qubits and process them on the same photonic chip \cite{Tagliavacche2025,10.1117/1.AP.6.3.036003}. Alternatively, by incorporating microresonator arrays \cite{PRXQuantum.6.010338}, qudit states can be generated. 
Although frequency-bin sources offer inherently low SWaP requirements, the free-space analyzer tends to become increasingly bulky as the state dimensionality grows. 
A possible solution lies in the use of Herriott cells \cite{Herriott:65} or nested interferometers \cite{PhysRevLett.109.023601}, which allow the implementation of unbalanced interferometers with significantly reduced physical footprint, making them suitable for compact space-based photonic systems. 
 
 Moreover, frequency-bin encoding can serve as a versatile optical interface between heterogeneous network nodes \cite{Lingaraju:22}. 
     The narrow bandwidth of frequency-bin photons enables efficient storage and retrieval in quantum memories \cite{Wang:08}. 
 The entangled photon source and field-widened interferometers are both also compatible with time-bin encoding, making them suitable platforms for exploring time-frequency entanglement and hybrid encoding schemes. 
 For example, single-photon frequency combs, or time-frequency Gottesman-Kitaev-Preskill (TFGKP) states \cite{PhysRevLett.130.200602}, have been proposed to realize universal linear optical quantum computation (LOQC) without the use of active devices, in fact leveraging the exact building blocks demonstrated in this work: optical interleavers and time-resolved detectors. 
 The discretization in both time and frequency degrees of freedom leads to inherent error robustness against both temporal and spectral errors.

Although measurement fidelity is ultimately limited by the timing jitter of the detection system \cite{Vinet:25}, single-photon detector technologies have achieved timing resolutions as low as 3 ps FWHM in state-of-the-art demonstrations \cite{korzh_demonstration_2020} and 20-30 ps jitter in commercially available detectors \cite{esmaeil_zadeh_superconducting_2021}. Moreover, multi-mode fiber-coupled SNSPDs with $<20$ ps FWHM timing resolution have been demonstrated \cite{Chang:19}, indicating that such detectors can combine high temporal precision with efficient multi-mode collection thereby mitigating this constraint.

Our results present a simple and robust approach to manipulating frequency-bin qubits, paving the way toward the deployment of practical, scalable frequency-domain quantum information processing and communication systems.


\section*{Methods}\label{sec:methods}
\subsubsection*{Photonic chip characterization}
The photonic chip consists of a pair of Si$_3$N$_4$ microring resonators (R0 and R1), coupled to a single bus waveguide, with a cross-section $1.4\times0.4$ $\mu\text{m}^2$. Integrated resistive micro-heaters overlay both resonators, allowing for a fine-tuning of the spectral modes' distance via thermo-optic effect. The structure is embedded in a silica cladding. The device was designed and fabricated following the LIGENTEC AN400 multi-project-wafer run specifications. Light is coupled in and out from the chip with Ultra High Numerical Aperture (UNHA4) fiber array, ensuring $3$ dB/facet losses and mechanical stability. 

Figure~\ref{fig:timeresolved}\textcolor{blue}{c} shows the transmission spectra of R0 and R1 at signal, pump and idler frequencies. The loaded $Q$ of the resonances are $1.21\times10^6$ ($\ket{0_s}$), $1.23\times10^6$ ($\ket{1_s}$), $1.39\times10^6$ ($\ket{0_i}$) and $1.31\times10^6$ ($\ket{1_i}$). These results, combined with resonances' extinction, allow for an estimate intrinsic $Q_{int} \sim 3.5\times10^6$, propagation losses of $0.1$ dB/cm and escape efficiencies $\eta_e = \left(Q_{int}-Q\right) / Q_{int}$ ranging from $0.603$ to $0.655$. Transmission spectrum shows split pump resonances. This phenomenon is caused by Rayleigh backscattering, related to side-wall roughness, in high-$Q$ resonators: frontward and backward propagating modes interfere destructively, splitting single-peak resonances to two-peak resonances. A complete analysis of microring resonator backscattering can be found in Ref~\cite{Matres:17}.  

We estimate the microrings spectral brightness to be between \SI[per-mode=symbol]{5e5}{cps\per\milli\watt^{2}\per\giga\hertz} and \SI[per-mode=symbol]{5.5e5}{cps\per\milli\watt^{2}\per\giga\hertz}. The coincidence-to-accidentals ratio (CAR$\sim 4$) is mainly limited by spontaneous Raman scattering. Raman scattering arises in silicon nitride and silica on-chip, and in the UNHA4 fiber, due to the high germanium doping concentration. The detrimental effects of Raman scattering can be mitigated by employing narrowband filters after the chip, thereby dropping accidental counts, or by optimizing the device to increase mode confinement within the silicon nitride waveguide.

\subsubsection*{Interferometric design}
We built two symmetric interferometers based on the design shown in Fig.~\ref{fig:ZZmeasurement}\textcolor{blue}{b}. The signal and idler f-bin analyzers differ only in the input fiber: single-mode for the idler and multi-mode for the signal.
To compensate for the different spatial modes, we utilized a $10$ cm N-BK7 glass rod (\textit{Edmund Optics}) with a $1.5007$ refractive index at $1550$ nm. In the long arm of the interferometer, the beam propagated in a double-pass configuration through $1$cm of air from the beam splitter to the glass rod, then through the $10$ cm rod, followed by another $1$ cm air gap between the rod and the corner cube retro-reflector (CCR) for an optical path length of $34$ cm. The short arm, in comparison, had a physical length of $7.85$ cm, resulting in an optical path length difference of $\Delta L\sim 18.3$ cm corresponding to a \SI{609.8}{\pico\second} time delay or equivalently a $\Delta\omega/2\pi = \SI{820}{\mega\hertz}$ frequency splitting. The glass rod slightly perturbed the polarization state of the transmitted light. To characterize its effect, we performed a polarization-extinction measurement. Inserting the rod between two orthogonal linear polarizers reduced the extinction ratio from $43$ dB to $20$ dB, likely due to stress-induced anisotropy in the glass. Consistent with this hypothesis, loosening the rod in its mount improved the extinction to $29$dB.

To carefully match the optical path length difference in both interferometers we operated the \textit{Santec TSL-570} in continuous linear scan mode. The laser output was injected through both MZI sequentially, and the interference signal at the output was monitored on a fast mixed-domain oscilloscope (\textit{Tektronix MDO3034}). As the laser frequency was swept, each interferometer produced a periodic transmission pattern corresponding to its individual free spectral range (FSR), determined by its path length difference $\Delta L$. Consequently, by suppressing the resulting beat note from mismatched FSRs, we were able to carefully match the optical delays in both interferometers. 

Our field-widened interferometers can be further optimized. The throughput could be significantly improved by employing optics with anti-reflection (AR) coatings and using glass rods with higher surface quality, as the current rods contributed noticeably to the optical loss. We summarize the losses for each interferometer in Table~\ref{tab:losses}.
\begin{table*}
    \centering
    \begin{tabular}{|c|c|c|c|}
    \hline
        Channel & Interferometer loss (dB) &
        Detector efficiency ($\%$) & $\eta_T$  \\
          \hline
        Idler MZI Output 1 & 5.1 & 0.85 & 0.2635\\
         Idler MZI Output 2 & 5.3 & 0.85 & 0.2499\\
          Signal MZI Output 1 & 10.7 & 0.85 & 0.0723\\
           Signal MZI Output 2 & 5.4 & 0.15 & 0.0429\\
           \hline
    \end{tabular}
    \caption{Channel efficiencies ($\eta_T$) for the $\sigma_Z$-basis measurements. }
    \label{tab:losses}
\end{table*}
Furthermore, interferometer visibility can be enhanced through more precise alignment. Visibilities approaching unity for the multi-mode analyzers were achieved in Waterloo before shipment to Pavia. The improved throughput would significantly reduce the measurement acquisition time, thus alleviating the phase stability requirements for the interferometers. Additionally, improvements to the entangled photon source, particularly in mitigating Raman scattering in the SiN ring resonators, could boost the CAR, removing the need for background subtraction and enabling higher-fidelity measurements. 
\subsubsection*{Multi-mode characteristics}
We characterize the multi-mode nature of the beam profile imaged with a CCD camera in the signal interferometer. To evaluate the modal expansion, we filter the point spread function (PSF) to retain only the dominant spatial frequency components. We then reconstruct the PSF from its modal decomposition over the first 30 Hermite-Gaussian modes (see Supplementary Note 3). 
This modal spectrum is typical of an urban-scale free-space link attesting to the validity of the proposed measurement scheme from the perspective of a real-world scenario \cite{Vinet:25,10.1117/1.OE.57.7.076106}.

\section*{Data availability}
The analysis code is provided in Ref.~\cite{vinet2025frequencybin}.
The data sets used and analyzed in the current study are available from the corresponding author on reasonable request.
\begingroup
\scriptsize	
\bibliography{References}
\endgroup
\section*{Acknowledgements}

{\small 
The authors thank Micro Photon Devices for kindly providing the MPD PDM-IR detector. The authors thank Kaylee Sherk, Kimia Mohammadi, Paul Godin, Katanya Kuntz and Giuseppe Vallone for helpful discussions. 
}
\section*{Funding}
This work is co-funded by the Natural Sciences and Engineering Research Council of Canada (NSERC) and the European Union (EU) under Grant no. 101070168 (HyperSpace).  The authors further acknowledge support from the NRC Quantum Sensing Challenge Program (QSP-019), the Canada Excellence Research Chair program (CERC), the Canadian Foundation for Innovation (CFI), the Ontario Research Fund (ORF) and the Institute for Quantum Computing. S.V. thanks the NSERC CGS-D for personal funding.
M.C. and M.B acknowledge support by the Italian Ministry of Education (MUR) PNRR project PE0000023-NQSTI.
\section*{Author contributions}
S.V., Y.Z., M.B., M.Ga., D.B. and T.J. conceived the original idea. 
M.C. and Y.Z. developed the theoretical framework.
M.C. and M.B. developed the frequency-bin entangled photon source. 
S.V. and M. Gi. engineered and fabricated the field-widened interferometers with the assistance of L.N.
S.V., M.C., M.B. and M. Gi. performed the experimental measurements. 
P.V., M.Ga., D.B. and T.J. supervised the experiment. 
S.V. performed the data analysis with the assistance of M.C., Y.Z. and M.B. 
S.V., M.C. and Y.Z. wrote and revised the manuscript. 
All authors commented on the manuscript.
{\small 

}

\section*{\protect Competing interests}

{\small The authors declare no competing interests.}

\section*{\protect Additional information}

{\small {\bf Supplementary information} Supplementary information is provided below.}

\noindent {\small {\bf Correspondence} and requests for materials should be addressed to St\'ephane Vinet.}
\section*{Supplementary Note 1 - Equatorial measurement}
\begin{figure*}
    \centering
    \includegraphics[width=0.7\linewidth]{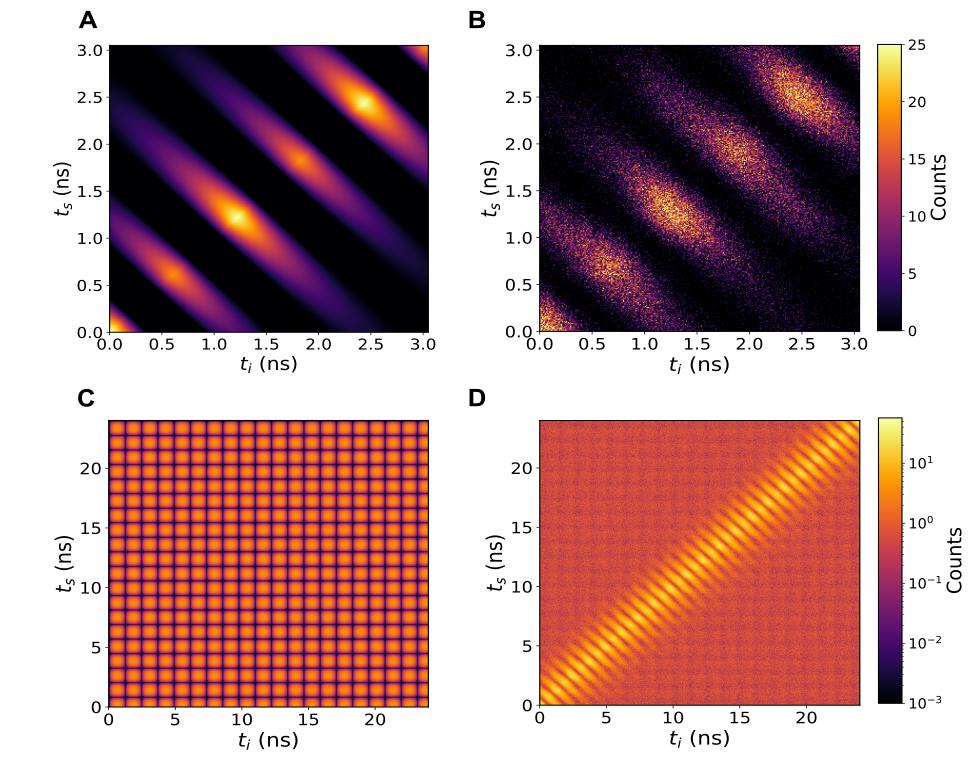}
    \caption{\small \textbf{ Joint temporal intensity.}
    \textbf{(a)} Theoretical JTI for three frequency components each separated by $\Delta\omega/2$.
    \textbf{(b)} Measured JTI over single-mode fiber.
    \textbf{(c)} JTI for a classical bichromatic pump.
    \textbf{(d)} Measured JTI with logarithmic color scale.
    For panels (c) and (d), the time axes have been rescaled from 3.05 ns to 24 ns. }
    \label{fig:qutrit}
\end{figure*}
 To understand the observed asymmetry between the $\langle\sigma_X\sigma_X\rangle$ and $\langle\sigma_Y\sigma_Y\rangle$ correlations, we simulate in Fig.~\ref{fig:qutrit}\textcolor{blue}{a} the joint temporal intensity (JTI) under the assumption that leakage of the bichromatic pump, enhanced by spontaneous Raman scattering, introduces weak incoherent sidebands at nearby frequencies. These components then interfere to produce an additional modulation at $\Delta\omega/2$ leading to an oscillatory term $\cos\!\bigl(\Delta\omega(t_\mathrm{s}+t_\mathrm{i})+\theta_1\bigr)+\cos\!\bigl(\Delta\omega/2(t_\mathrm{s}+t_\mathrm{i})+\theta_2\bigr)$ along the diagonal of the JTI,  
 consistent with the parity-dependent oscillations observed experimentally in Fig.~\ref{fig:qutrit}\textcolor{blue}{b}. For the bichromatic pump, $J(t_s,t_i)\propto \cos^2(\Delta\omega t_s)\cos^2(\Delta\omega t_i)$ results in a checkered pattern as shown in Fig.~\ref{fig:qutrit}\textcolor{blue}{c}. This pattern can be seen in the background of the experimental JTI in Fig.~\ref{fig:qutrit}\textcolor{blue}{d} indicating the presence of uncorrelated pump photons.
\subsection*{Phase tuning}
The phase of the equatorial basis measurement can be adjusted either by tuning the joint detection time $t_s+t_i$ or by varying the relative arrival time of the signal and idler photons with respect to the phase modulation signal.
The correlation fringes for the latter approach, corresponding to different $\Delta t=t_\mathrm{s}-t_\mathrm{i}$, are plotted in Fig.~\ref{fig:offdiag}. Note that there is a drop in fringe visibility with increasing $\Delta t$ due to an under-correction of the decay induced by the finite cavity lifetime.
\subsection*{Scalability to higher dimensions}
\begin{figure*}
     \centering
     \includegraphics[width=\linewidth]{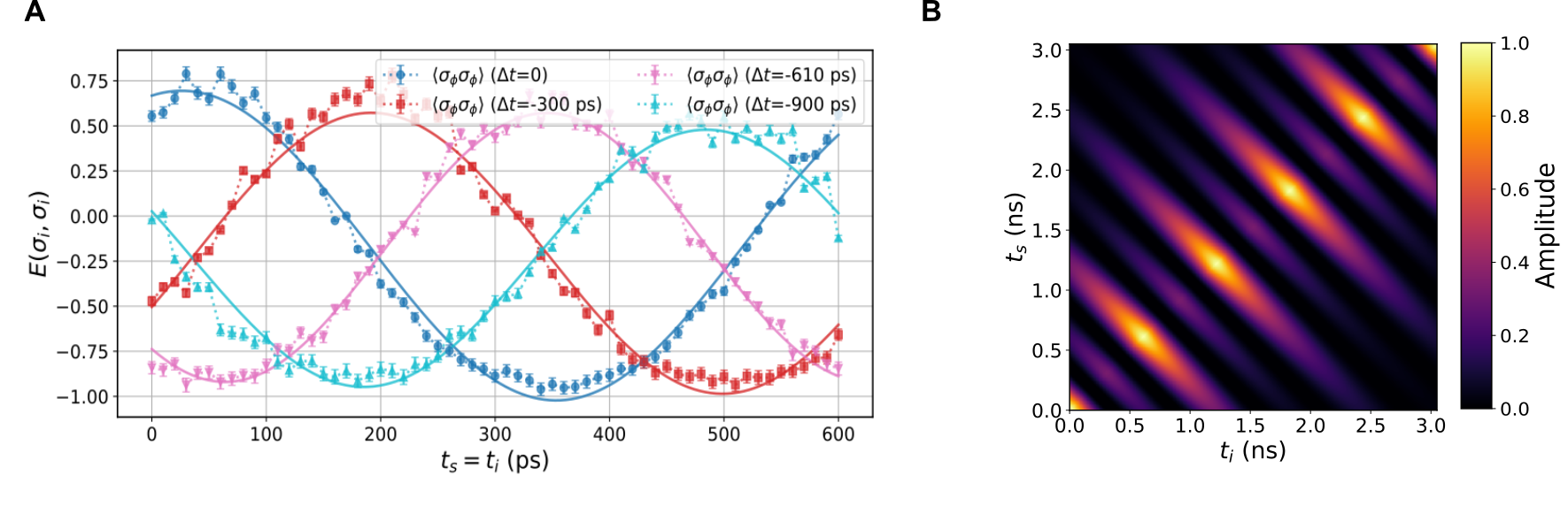}
      \caption{\small \textbf{ $\langle \sigma_\phi\sigma_\phi \rangle$.}  \textbf{(a)} Correlation function $\langle \sigma_\phi\sigma_\phi \rangle$ as a function of the absolute time $\tau$ for different relative time delays $\Delta t=t_s-t_i$. Here $\phi$ denotes the azimuthal angle on the equator of the Poincar\'e sphere. Due to the cavity lifetime,  the amplitude of the oscillation decays proportional to $|t_s-t_i|$ suggesting under-corrections in our time-dependent detector modeling.  \textbf{(b)} JTI simulation for the time-resolved detection of a qutrit as described in Eq.~\ref{eq:qutrit}.}
     \label{fig:offdiag}
 \end{figure*}

Time-resolved detection provides access to many measurements bases in higher dimensions as discrete frequency-bins produce interference fringes in the joint temporal intensity. For instance, a simulation corresponding to the qutrit projection given in Eq.~\ref{eq:qutrit} is presented in Fig.~\ref{fig:offdiag}\textcolor{blue}{b}. 
    \begin{multline}\label{eq:qutrit}
        \ket{P_{s,i}(t_s,t_i)}\propto \\ \bigl(\ket{0_s}+e^{i\Delta\omega t_s}\ket{1_s}+e^{i2\Delta\omega t_s}\ket{2_s}\bigr)\otimes \bigl(\ket{0_i}+e^{i\Delta\omega t_i}\ket{1_i}+e^{i2\Delta\omega t_i}\ket{2_i}\bigr).\end{multline}
\section*{Supplementary Note 2 - Interferometric measurements}
\begin{figure*}[!t]
    \centering
    \includegraphics[width=\linewidth]{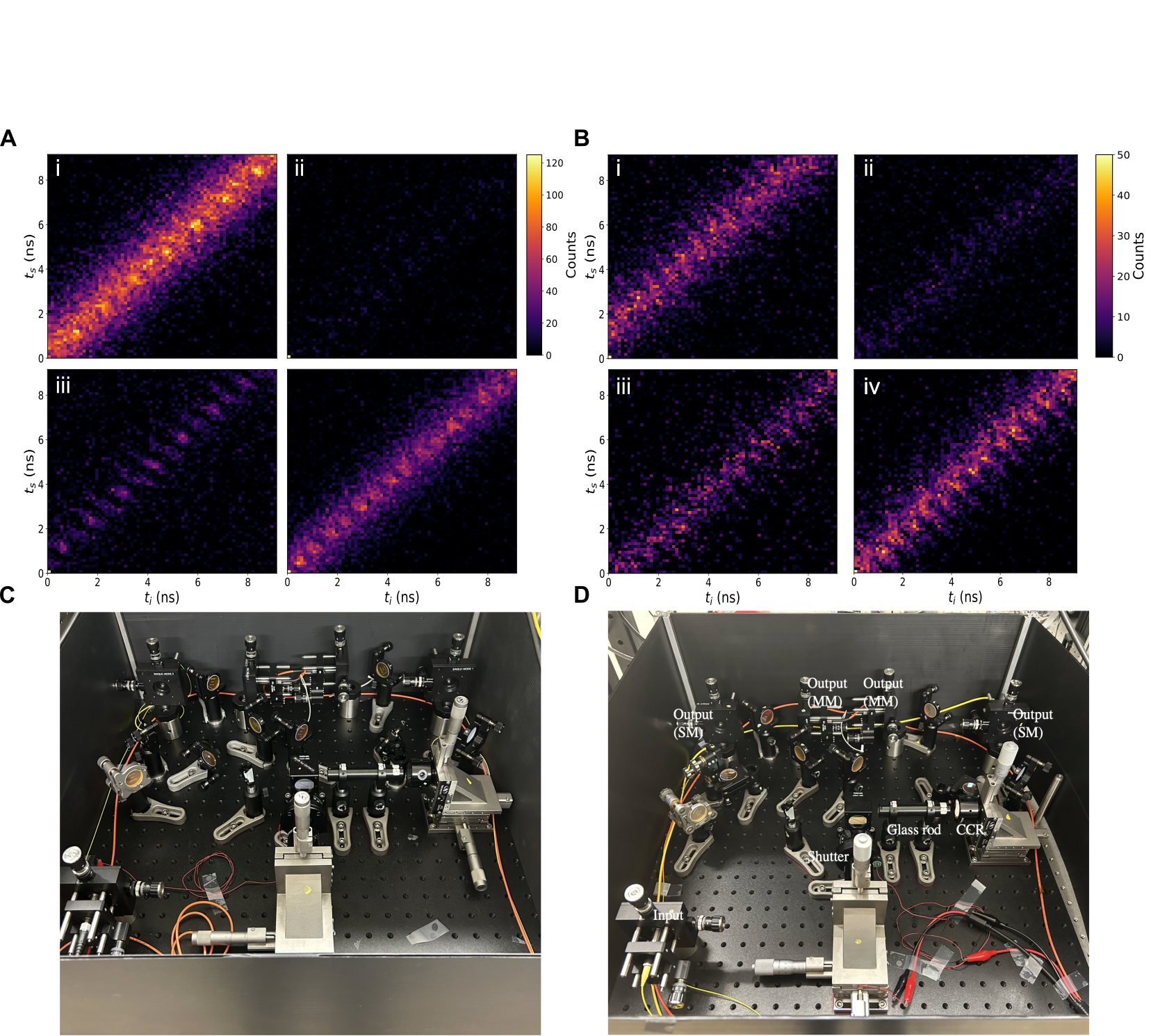}
    \caption{\small \textbf{ $\sigma_Z-$ basis measurements} 
    \textbf{(a)} $\sigma_Z\sigma_Z$ basis: experimental JTIs in i-iv correspond to the projections $\ketbra{00}{00}, \ketbra{01}{01},\ketbra{10}{10},\ketbra{11}{11}$ respectively. \textbf{(b)} i-ii experimental JTIs corresponding to corresponding to equatorial-$\sigma_Z$ measurements, iii-iv JTIs $\sigma_Z-$equatorial basis measurements. Photographs of the \textbf{(c)} idler photon's frequency-bin analyzer : the FWI has a single-mode fiber input, \textbf{(d)} the signal photon's frequency-bin analyzer: the FWI has a multi-mode fiber input (CCR: corner cube retroreflector).  }
    \label{fig:JTIZZ}
\end{figure*}

\begin{figure}
     \centering
     \includegraphics[width=\linewidth]{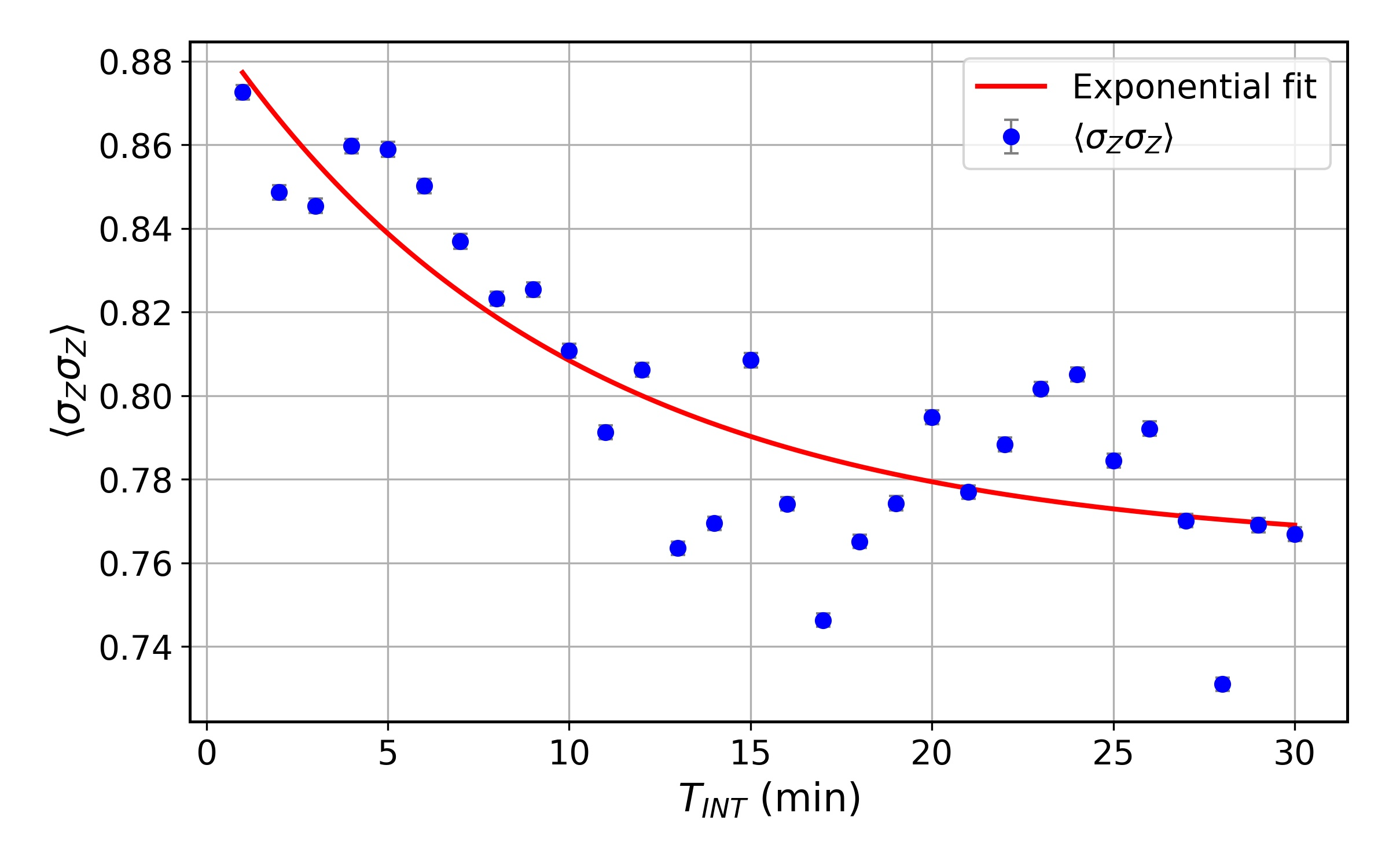}
      \caption{\small \textbf{ $\langle\sigma_Z\sigma_Z\rangle(t)$ :}  phase fluctuations in both interferometers lead to an exponential decay of $\langle\sigma_Z\sigma_Z\rangle$. }
     \label{fig:ZZtimeerror}
 \end{figure}

\noindent
The JTIs for the $\sigma_Z\sigma_Z$ and cross-basis measurements, where one photon is measured in the $\sigma_Z-$basis and the other in the equatorial-basis are shown in Fig.~\ref{fig:JTIZZ}. For the $\sigma_Z\sigma_Z$-basis in Fig.~\ref{fig:JTIZZ}\textcolor{blue}{a}, the four JTIs correspond to the $\ketbra{ij}{ij}$ projections, where $ i,j\in\{0,1\}$.  To be tomographically complete, we must also consider the cross-basis measurements where one photon is measured in the $\sigma_Z-$ basis and the other in the equatorial basis. We present the JTIs associated to the signal in the equatorial basis and idler in the $\sigma_Z$ basis followed by the signal in the $\sigma_Z$ basis and idler in the equatorial basis in Fig.~\ref{fig:JTIZZ}\textcolor{blue}{b i-iv}. By post-selecting on detection times $t_i$ and $t_s$ we can implement projections in the $\sigma_Z\sigma_X, \sigma_Z\sigma_Y, \sigma_X\sigma_Z,\sigma_Y\sigma_Z$ bases as required for quantum state tomography.  
In our implementation, both interferometers (pictured in Fig.~\ref{fig:JTIZZ}\textcolor{blue}{c-d}) operated without active phase stabilization, which constrained the measurement integration time $T_{INT}$. Indeed, the phase in each interferometer drifts randomly according to environmental fluctuations. If both phase shifts follow a random walk, the variance of the interferometers' relative phase $\Delta\phi(t)$ grows linearly over time which leads to an exponential decay of $\langle\sigma_Z\sigma_Z\rangle$ as observed in Fig.~\ref{fig:ZZtimeerror}. Specifically, $\langle\sigma_Z\sigma_Z\rangle(t)$ decays proportionally to $\langle\sigma_Z\sigma_Z\rangle(t=0)\times e^{-2Dt},$ where $D$ is the diffusion constant reflecting the phase noise introduced by the environment.  
\section*{Supplementary Note 3 - Multi-mode characteristics}
\begin{figure}
    \centering
    \includegraphics[width=\linewidth]{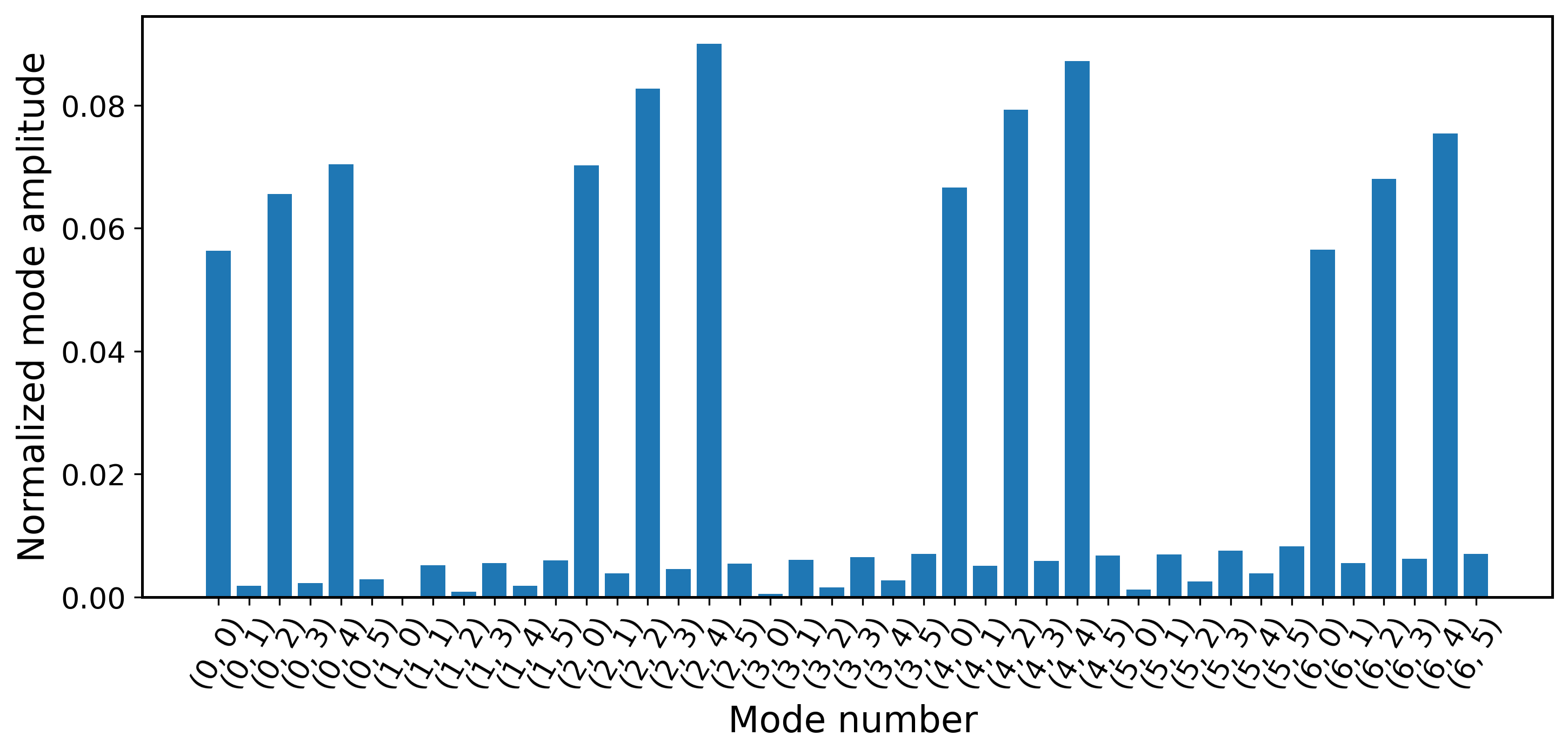}
    \caption{\small{}  Modal decomposition of the point spread function into the first 30 Hermite-Gaussian modes $(n,m)$.}
    \label{fig:modedecomp}
\end{figure}

\noindent
To characterize the multi-mode channel we image the beam profile in the signal interferometer with a CCD camera (\textit{New Imaging Technologies WiDy SenS 640}) and perform a modal decomposition of the point spread function (PSF). The PSF is expanded into the Hermite-Gaussian (HG) basis and the corresponding mode weights, in Fig.~\ref{fig:modedecomp}, provide a measure of how the optical power distribution among the first 30 spatial modes. From these weights, we extract an effective mode number $M\sim12$ indicating that the beam energy is predominantly shared among the twelve modes with a clear dominance of even-order HG modes consistent with the spatial symmetry of the measured PSF.
\end{document}